\documentstyle[preprint,aps,eqsecnum]{revtex}
\tightenlines
\def\m@thcombine#1#2{%
  \setbox0=\hbox{$#1$}
  \setbox1=\hbox{$#2$} 
  \ifdim\wd0>\wd1
    \setbox0=\hbox to\wd1{\hss\box0\hss}
  \else
    \setbox1=\hbox to\wd0{\hss\box1\hss}
  \fi
  \mathop{\vcenter{
    \offinterlineskip\box0\box1}}}
\def\lesim{\m@thcombine<\sim}
\def\gesim{\m@thcombine>\sim}

\begin{document}

\draft
\title{ HOW TO CALCULATE THE QUANTUM PART OF THE TRULY NONPERTURBATIVE        
YANG-MILLS VACUUM ENERGY DENSITY IN THE AXIAL GAUGE QCD}

\author{V. Gogohia and Gy. Kluge}

\address{HAS, CRIP, RMKI, Theory Division, P.O.B. 49, H-1525 Budapest 114, Hungary \\ email address: gogohia@rmki.kfki.hu }                                   

\maketitle

\begin{abstract}
Using the effective potential approach for composite operators, we have formulated a general method how to calculate the truly nonperturbative vacuum energy density in the axial gauge QCD quantum models of its ground state. It is defined
as integrated out the truly nonperturbative part of the full gluon propagator over the deep infrared region (soft momentum region). The nontrivial minimization procedure  
makes it possible to determine the value of the soft cutoff in terms of the corresponding nonperturbative scale parameter which is inevitably presented in any
nonperturbative model for the full gluon propagator. If the chosen Ansatz for the full gluon propagator is a realistic one, then our method uniquely determines the truly vacuum energy density, which is always finite, automatically negative
and it has no imaginary part (stable vacuum). We illustrate it by considering the Abelian Higgs model of dual QCD ground state. We have explicitly shown that 
the vacuum of this model without string contributions is unstable against quantum corrections. 
\end{abstract}


\vfill

\eject

\section{Introduction}

The nonperturbative QCD vacuum is a very complicated medium and its dynamical 
and topological complexity [1-3] means that its structure can be organized at various 
levels (classical, quantum) and it can contain many different components and ingredients which contribute to the vacuum energy density (VED), one of the main 
characterics of the QCD ground state.                                          
Many models of the QCD vacuum involve some extra classical color field 
configurations such as randomly oriented domains of constant color magnetic 
fields, background gauge fields, averaged over spin and color, stochastic colored background fields, etc. (see Refs. [1,4] and references therein).
The most elaborated classical models are random and interacting instanton liquid models (RILM and IILM, respectively) of the QCD vacuum [5]. These models are 
based on the existence of the topologically nontrivial instanton-type fluctuations of gluon fields, which are nonperturbative solutions to the classical equations of motion in Euclidean space (see Ref. [5] and references therein).

Here we are going to discuss the quantum part of VED which is determined by the
effective potential approach for composite operators 
[6,7] (see also Ref. [8]). It allows us to investigate the nonperturbative QCD
vacuum, in particular Yang-Mills (YM) one, by substituting some physically     
well-justified Ansatz for the full gluon propagator since the exact solutions are not known. In the absence of external sources the effective potential is nothing but VED which is given in the form of the loop expansion where the number 
of the vacuum loops (consisting in general of the confining quarks and nonperturbative gluons) is equal to the power of the Plank constant, $\hbar$.          
              
Let us remind the reader that the full dynamical information of any quantum gauge field theory such as QCD is contained in the corresponding quantum equations
of motion, 
the so-called Schwinger-Dyson (SD) equations for lower (propagators) and higher
(vertices and kernels) Green's functions [9,10]. These equations should be also
complemented
by the corresponding Slavnov-Taylor (ST) identites [9-12] which in general relate the above mentioned lower and higher Green's functions to each other.       
These identities are consequences of the exact gauge invariance and therefore
$"are \ exact \ constaints \ on \ any \ solution \ to \ QCD"$ [9]. Precisely this system of equations can serve as an adequate and effective tool for the nonperturbative approach to QCD. Among the above-mentioned Green's functions, the two-point Green's function describing the full gluon propagator (see section II 
below) has a central place [9-13]. In particular, the solutions to the above-mentioned SD equation for the full gluon propagator, are supposed to reflect the 
quantum structure of the QCD ground state. It is a highly nonlinear integral equation containing
many different propagators, vertices and kernels [9-13]. For this reason it may
have many different exact solutions with different asymptotics in the deep infrared (IR) limit (the ultraviolet
(UV) asymptotics because of asymptotic freedom are apparently uniquely determined), describing thus many different types of quantum excitations of gluon field
configurations in the QCD vacuum. Evidently, there is no hope for an exact solutions as well as not all of them can reflect the real structure of the QCD vacuum. Let
us emphasize now that any deviation in the behavior of 
the full gluon propagator in the IR domain from the free one automatically
assumes its dependence on a scale parameter (at least one) resposible for nonperturbative dynamcs in the quntum model under consideration, say, $\Lambda_{NP}$. This is very similar to asymtotic freedom which requires  asymptotic scale
parameter associated with the nontrivial perturbative dynamics (scale violation). However, to calculate the truly nonperturbative VED we need not the IR part 
in the decomposition of the full gluon propagator, but rather
its truly nonperturbative part which
wanishes when the above-mentioned nonperturbative scale parameter goes to
zero, i.e., when the perturbative phase survives only in the corresponding decomposition of the full gluon propagator (see next section below). 

 It is well known, however, that VED is badly divergent in quantum field       
theory, in particular QCD (see, for example, the discussion given by Shifman in
Ref. [1]). The main problem
thus is how to extract the truly nonperturbative VED which is relevant for the 
QCD vacuum quantum model under consideration. It should be finite, negative and
it should have no imaginary part (stable vacuum).
Why is it so important to calculate it from first principles? As was emphasized
above, this quantity is important in its own right being nothing but the
bag constant (the so-called bag pressure) apart from the sign, by definition [14]. Through the trace anomaly relation [15] it assists in the correct estimating such an important
phenomenological nonperturbative parameter as the gluon condensate introduced in the QCD sum rules approach to resonance physics [16]. Furthermore, it assists
in the resolution of the $U(1)$ problem [17] via the Witten-Veneziano (WV) formula for the mass of $\eta'$ meson [18]. The problem is that the topological susceptibility needed for this purpose [16-19] is determined by the two point correlation function from which perturbative contribution
is already subtracted by definition [18-22]. The same is valid for the above-mentioned bag constant which is much more general quantity than the string tension since it is relevant for light quarks as well. Thus to correctly calculate the truly nonperturbative VED means to correctly understand 
the structure of the QCD vacuum in different models.                           
    
We have already formulate a method how to calculate the truly nonperturbative YM VED in the covariant gauge QCD [23]. The main purpose of this paper (section 
II) is to formulate precisely a general method how to correctly calculate the truly nonperturbative quantum part of YM VED in the axial gauge QCD. In sections
III and IV we illustrate it
by considering the Abelian Higgs model [24] of the dual QCD [25] ground state. 
We will explicitly show that the vacuum of this model without string contributions is unstable against quantum corrections. In section V we summarize our results.

\section{The truly nonperturbative vacuum energy density }

 In this section we are going to analytically formulate a general method of calculation of the quantum part of the truly nonperturbative YM VED in the axial gauge QCD.
Let us start from the nonperturbative gluon part of VED
which to-leading order (log-loop level $\sim \hbar$)\footnote{Next-to-leading and higher terms (two and more vacuum loops) are suppressed 
by one order of magnitude in powers of $\hbar$ at least and are left for consideration elsewhere.} 
is given by the effective potential for composite operators [6] as follows

\begin{equation}
V(D) =  { i \over 2} \int {d^nq \over {(2\pi)^n}}
 Tr\{ \ln (D_0^{-1}D) - (D_0^{-1}D) + 1 \}, 
\end{equation}
where $D(q)$ is the full gluon propagator (see below) and $D_0(q)$ is its  
free (perturbative) counterpart. Here and below the traces over
space-time and color group indices are understood.
The effective potential is normalized as $V(D_0) = 0$, i.e., the free perturbative vacuum is normalized to zero.                                              

A general parametrization of the gauge boson propagator in the axial gauge of dual QCD is [24-26] (here and below we use notations and definitions of Refs. [24,26])

\begin{equation}
D_{\mu\nu}(q, n) = - { 1 \over (q \cdot n)^2} T_{\mu\nu} (n) G(-q^2) + L_{\mu\nu}(q, n) F(-q^2),
\end{equation}
where

\begin{eqnarray}
T_{\mu\nu}(n) &=& \delta_{\mu\nu} - n_{\mu} n_{\nu}, \nonumber\\               
L_{\mu\nu}(q, n) &=& \delta_{\mu\nu} - { q_{\mu} n_{\nu} + q_{\nu} n_{\mu}
\over (q \cdot n)} + { q_{\mu} q_{\nu} \over (q \cdot n)^2}
\end{eqnarray}
with an arbitrary constant unit vector $n_{\mu}$, $n^2_{\mu} =1$. The exact coefficient functions $G(-q^2)$ and $F(-q^2)$ characterize the vacuum of the theory under consideraion. Their free perturbative counterparts are

\begin{equation}
F^{PT} (-q^2) = { 1 \over (- q^2)}, \qquad  G^{PT}(-q^2) = 0. 
\end{equation}
Thus the free perturbative gluon propagator is

\begin{equation}
D^0_{\mu\nu}(q, n) = { 1 \over (-q^2)} L_{\mu\nu}(q, n)
\end{equation}
while its inverse is

\begin{equation}
[D^0_{\mu\nu}]^{-1}(q) = (-q^2) \left( \delta_{\mu\nu} - { q_{\mu} q_{\nu} \over q^2 } \right). 
\end{equation}
Using futher Eqs. (2.2) and (2.6), one obtains

\begin{equation}
[D^0_{\mu\nu}]^{-1}(q)D_{\mu\nu}(q, n) = (-q^2) F(-q^2) + G(-q^2).
\end{equation}

In order to evaluate the
effective potential (2.1) we use the well-known expression,

\begin{equation}
Tr \ln (D_0^{-1}D) = 8 \times \ln det (D_0^{-1}D) =
8 \times 4 \ln \left[ (-q^2)F(-q^2) + G(-q^2) \right].
\end{equation}
It becomes zero (in accordance with the above mentioned normalization condition) when the full gluon form factors are replaced by their free counterparts (see
Eqs. (2.4)). Going over to four ($n=4$) dimensional Euclidean space in Eq. (2.1), on account of (2.8), and evaluating some numerical factors, one obtains ($\epsilon_g = V(D)$)

\begin{equation}
\epsilon_g = - { 1 \over \pi^2} \int dq^2 \ q^2 
\left[ \ln \left( q^2 F(q^2) + G(q^2) \right) - \left( q^2 F(q^2) + G(q^2) \right) + 1 \right]. 
\end{equation}

Let us now introduce the following decomposition of the exact coefficient functions $G(q^2)$ and $F(q^2)$ (Euclidean metrics) 

\begin{eqnarray}
F(q^2) &=& F^{NP}(q^2) + F^{PT}(q^2), \nonumber\\
G(q^2) &=& G^{NP}(q^2) + G^{PT}(q^2),
\end{eqnarray}
where the truly nonperturbative quantities $F^{NP}(q^2)$ and $G^{NP}(q^2)$ are 
defined as follows: 

\begin{eqnarray}
F^{NP}(q^2, \Lambda_{NP}) &=& F(q^2, \Lambda_{NP})- F(q^2, \Lambda_{NP} = 0), \nonumber\\
G^{NP}(q^2, \Lambda_{NP}) &=& G(q^2, \Lambda_{NP})- G(q^2, \Lambda_{NP} = 0),
\end{eqnarray}
which explains the difference between the truly nonperturbative parts and the full gluon form factors which are nonperturbative themselves. Let us note, that 
the perturbative parts $F^{PT}(q^2)$ and $G^{PT}(q^2)$ may, in general, contain
renormgroup log improvmenets due to asymptotic freedom. Without these improvments their free perturbative counterparts are given in Eqs. (2.4). 
Substituting these relations into Eq. (2.9) and doing some trivial rearangment,
one obtains

\begin{equation}
\epsilon_g = - { 1 \over \pi^2} \int dq^2 \ q^2 
\left[ \ln \left(1 + q^2 F^{NP}(q^2) + G^{NP}(q^2) \right) - \left( q^2 F^{NP}(q^2) + G^{NP}(q^2) \right) \right] + I_{PT},
\end{equation}
where we introduce the following notation

\begin{equation}
I_{PT} = - { 1 \over \pi^2} \int dq^2 \ q^2 
\left[ \ln \left(1 - {1 - q^2 F^{PT}(q^2) - G^{PT}(q^2) \over 1 + q^2 F^{NP}(q^2) + G^{NP}(q^2)} \right) + \left( 1 - q^2 F^{PT}(q^2) - G^{PT}(q^2) \right) \right],
\end{equation}
as containing contribution which is mainly determined by the perturbative 
part. However, this is not the whole story yet. We must now to introduse the soft cutoff in order to separate the deep IR region where the truly nonperturbative contributions become dominant (obviously they can not be valid in the whole energy-momentum range). So the expression (2.12) becomes                       
      
\begin{equation}
\epsilon_g = - { 1 \over \pi^2} \int_0^{q_0^2} dq^2 \ q^2 
\left[ \ln \left(1 + q^2 F^{NP}(q^2) + G^{NP}(q^2) \right) - \left( q^2 F^{NP}(q^2) + G^{NP}(q^2) \right) \right] + I_{PT} + \tilde{I}_{PT}, 
\end{equation}
where the explicit formula for $\tilde{I}_{PT}$ (which is obvious) is not important. The contribution over the perturbative region $\tilde{I}_{PT}$ as well as
$I_{PT}$ should be subtracted by
introducing  the corresponding counter terms into the effective potential, which is equivalent to define the truly nonperturbative VED as
$\epsilon_g^{np} = \epsilon_g - I_{PT} - \tilde{I}_{PT}$. Thus one finally obtains 

\begin{equation}
\epsilon_g^{np} = - { 1 \over \pi^2} \int_0^{q_0^2} dq^2 \ q^2 
\left[ \ln \left(1 + q^2 F^{NP}(q^2) + G^{NP}(q^2) \right) - \left( q^2 F^{NP}(q^2) + G^{NP}(q^2) \right) \right]. 
\end{equation}
This a general formula which can be applied to any model of the axial gauge QCD
ground state based on the corresponding Ansatz for the full gluon propagator.  
So Eq. (2.15) is our definition of the truly nonperturbative
VED as integrated out the truly nonperturbative part of the full gluon propagator over the deep IR region, soft momentum region, $0 \leq q^2
\leq q^2_0$. How to determine $q_0^2$? By the corresponding minimization procedure, of course (see below).

\subsection{}

From this point it is convenient to factorize the dependence on a scale in the 
nonperturbative VED (2.15). As was already emphasized above, 
the full gluon form factors always
contain at least one scale parameter responsible for the nonperturbative dynamics in the model under consideration, $\Lambda_{NP}$. Within our general method 
we are considering it as free one, i.e., as "running" (when it formally goes 
to zero, only perturbative phase survives in the model under consideration) and
its numerical value (if any) will be used only at final stage in order to numerically evaluate the corresponding truly nonperturbative VED (if any).
We can introduce dimensionless variables and parameters by
using completely extra scale (which is aways fixed in comparison with $\Lambda_{NP}$), for example flavorless QCD asymptotic scale parameter $\Lambda_{YM}$ as
follows:                                                

\begin{equation}
z = {q^2 \over \Lambda_{YM}^2}, \quad z_0 = {q_0^2 \over \Lambda_{YM}^2}, 
\quad b= {\Lambda^2_{NP} \over \Lambda^2_{YM}}.                        
\end{equation}
Here $z_0$ is the corresponding dimensionless soft cutoff while the parameter  
$b$ has a very clear physical
meaning. It measures the ratio between nonperturbative dynamics, symbolized by 
$\Lambda^2_{NP}$ and nontrivial perturbative dynamics (violation of scale, asymptotic freedom) symbolized by $\Lambda^2_{YM}$. When it is zero only perturbative 
phase remains in the quantum model under consideration. In this case, the gluon
form factors obviously become a functions of $z$ and $b$, i.e.,  $F^{NP}(q^2)=F^{NP}(z, b)$ and  $G^{NP}(q^2)=G^{NP}(z, b)$, so the truly nonperturbative VED 
(2.15) is ($\epsilon_g^{np} \equiv \epsilon_g^{np}(z_0, b)$)

\begin{equation}
\Omega_g (z_0, b) = { 1 \over \Lambda_{YM}^4} \epsilon_g^{np}(z_0, b),
\end{equation}
where for futher aims we introduce the gluon effective potential at a fixed scale $\Lambda_{YM}$, [23,27]                                                  
    
\begin{equation}
\Omega_g \equiv \Omega_g (z_0, b) = {1 \over \pi^2}   
\int_0^{z_0} dz \ z \left[ \left(zF^{NP}(z,b) + G^{NP}(z,b) \right) -  \ln \left( 1 + zF^{NP}(z,b) + G^{NP}(z,b) \right) \right].
\end{equation}
Precisely this expression allows us to investigate the dynamical structure of the YM vacuum free of scale dependence complications as it has been already factorized in Eq. (2.17). It depends only on $z_0$ and $b$ and the minimization procedure can be done now with respect to $b$,                                    
$ \partial \Omega_g (z_0, b) / \partial b = 0$
(usually after integrated out in Eq.
(2.18)) in order to find self-consistent relation between $z_0$ and $b$, which 
means to find $q_0$ as a function of $\Lambda_{NP}$.                           
 Let us note in advance that all final numerical results will always depend only on $\Lambda_{NP}$ as it should be for the nonperturbative part of VED.
Obviously, the minimization with respect to $z_0$ leads to trivial zero. In principle, through the relation $\Lambda_{YM}^4 = q_0^4 z_0^{-2}$, it is possible 
to fix the soft cutoff $q_0$ itself, but this is not the case indeed since then
$z_0$ can not be varied.

\subsection{}

  On the other hand, the scale dependence can be factorized as follows: 

\begin{equation}
z={q^2 \over \Lambda^2_{NP}}, \quad  z_0={q_0^2 \over \Lambda^2_{NP}},     
\end{equation}
i.e., $b=1$. For simplicity (but not loosing generality) we use the same notations for the dimensionless set of variables and parameters as in Eq. (2.16). In 
this case, the gluon
form factors obviously becomes the function of $z$ only and the truly nonperturbative VED (2.15) becomes

\begin{equation}
\epsilon^{np}_g (z_0) = {1 \over \pi^2} q_0^4 z_0^{-2}
\int_0^{z_0} dz \ z \left[ \left(zF^{NP}(z) + G^{NP}(z) \right) -  \ln \left( 1 + zF^{NP}(z) + G^{NP}(z) \right) \right].
\end{equation}
Evidently, to fix the scale now is possible in the two different ways. In principle, we can fix $\Lambda_{NP}$ itself, i.e., introducing

\begin{equation}
\tilde{\Omega}_g (z_0) = { 1 \over \Lambda^4_{NP}} \epsilon_g^{np}(z_0)=       
{1 \over \pi^2}
\int_0^{z_0} dz \ z \left[ \left(zF^{NP}(z) + G^{NP}(z) \right) -  \ln \left( 1 + zF^{NP}(z) + G^{NP}(z) \right) \right].
\end{equation}
However, the minimization procedure again leads to the trivial zero, which shows that this scale can not be fixed.

In contrast to the previous case, let us fix the soft cutoff itself, i.e., 
setting [23,28]

\begin{equation}
\bar \Omega_g (z_0) = { 1 \over q_0^4} \epsilon_g^{np
}(z_0)= {1 \over \pi^2}  z_0^{-2}                                              
\int_0^{z_0} dz \ z \left[ \left(zF^{NP}(z) + G^{NP}(z) \right) -  \ln \left( 1 + zF^{NP}(z) + G^{NP}(z) \right) \right].
\end{equation}
The minimization procedure with respect to $z_0$ is nontrivial now. Indeed,
$ \partial \bar \Omega_g (z_0) / \partial z_0 = 0$, yields the following "stationary" condition 

\begin{eqnarray}
\int_0^{z_0} dz \ z \left[ \left(zF^{NP}(z) + G^{NP}(z) \right) - \ln \left( 1 + zF^{NP}(z) + G^{NP}(z) \right) \right] \nonumber\\                        
= {1 \over 2} z_0^2 \left[ \left( z_0F^{NP}(z_0) + G^{NP}(z_0) \right) -  \ln \left( 1 + z_0F^{NP}(z_0) + G^{NP}(z_0) \right) \right],
\end{eqnarray}
which solutions (if any) allows one to find $q_0$ as a function of $\Lambda_{NP}$.
On account of this "stationary" condition, the effective potential (2.22) itself becomes simpler for numerical calculations, namely

\begin{equation}
\bar \Omega_g (z_0^{st}) = {1 \over 2 \pi^2} \left[ \left( z_0^{st} F^{NP}(z_0^{st}) + G^{NP}(z_0^{st}) \right) -  \ln \left( 1 + z_0^{st} F^{NP}(z_0^{st}) + G^{NP}(z_0^{st}) \right) \right],
\end{equation}
where $z_0^{st}$ is a solution (if any) of the "stationary" condition (2.23) and corresponds to the minimum(s) (if any) of the effective potential (2.22).
In the next sections we will illustrate how this method works.

\section{Abelian Higgs model}

Let us now consider some special model of the dual QCD [25] ground state. In the dual Abelian Higgs theory which confines electric charges the coefficient functions $F(q^2)$ and $G(q^2)$ are [24] (Euclidean metrics)  

\begin{eqnarray}
F(q^2) &=& { 1 \over q^2 + M^2_B } \left( 1 + { M^4_B D^{\Sigma}(q^2) \over q^2 + M^2_B} \right) , \nonumber\\                                                
G(q^2) &=& - { M^2_B \over q^2 + M^2_B } \left( 1 - M^2_B { q^2 D^{\Sigma}(q^2) \over q^2 + M^2_B} \right),
\end{eqnarray}
where $M_B$ is the mass of the dual gauge boson $B_{\mu}$ and $D^{\Sigma}(q^2)$
represents the string contribution into the gauge boson propagator. The mass scale paprameter $M_B$ is the scale responsible for nonperturbative dynamics in this model (in our notations $\Lambda_{NP} = M_B$). When it formally goes to zero, then one recovers the free perturbative expressions indeed, (2.4). Removing 
the string contributions from these relations we get

\begin{equation}
F^{no-str.}(q^2) = { 1 \over q^2 + M^2_B }, \qquad                      
G^{no-str.}(q^2) = - { M^2_B \over q^2 + M^2_B },
\end{equation}
i.e., even in this case these quantities remain nonperturbative. 
The truly nonperturbative expressions (2.11) now become

\begin{eqnarray}
F^{NP}(q^2) &=& - { M_B^2 \over q^2 (q^2 + M^2_B) } \left( 1 - { M^2_B q^2 D^{\Sigma}(q^2) \over q^2 + M^2_B} \right) , \nonumber\\                           
G^{NP}(q^2) &=& - { M^2_B \over q^2 + M^2_B } \left( 1 - { M^2_B q^2 D^{\Sigma}(q^2) \over q^2 + M^2_B} \right),
\end{eqnarray}
while with no-string contributions they are

\begin{equation}
F^{no-str.}_{NP}(q^2) = - { M^2_B \over q^2 (q^2 + M^2_B) }, \qquad     
G^{no-str.}_{NP}(q^2) = - { M^2_B \over q^2 + M^2_B }.
\end{equation}
Both expressions (3.3) and (3.4) are truly nonperturbative indeed, since they become zero in the perturbative limit ($M_B \longrightarrow 0$), when only perturbative phase remains. From these relations also follows

\begin{eqnarray}                           
G^{NP}(q^2) &=& q^2 F^{NP}(q^2) = - { M^2_B \over q^2 + M^2_B } \left( 1 - {M^2_B q^2 D^{\Sigma}(q^2) \over q^2 + M^2_B} \right), \nonumber\\
G^{no-str.}_{NP}(q^2) &=& q^2 F^{no-str.}_{NP}(q^2) = - { M^2_B \over q^2 + M^2_B },
\end{eqnarray}
so the truly nonperturbative vacuum energy density (2.15) will depend only on  
one function, say, $G^{NP}(q^2)$ (see next section).

Although the expression (2.2), on account of (3.1), for the gluon propagator is
exact, nevertheless it contains an unknown function $D^{\Sigma}(q^2)$ which is the intermadiate string state contribution into the gauge boson propagator [24]. It can be considered as a glueball state with the photon quantum numbers $1^-$. The bahavior of this function $D^{\Sigma}(q^2)$ in the IR region ($q^2 \rightarrow 0$) can be estimated as follows [24]: 

\begin{equation}                           
D^{\Sigma}(q^2) = {C \over q^2 + M^2_{gl}} + ..., 
\end{equation}
where $C$ is a dimensionless parameter and $M^2_{gl}$ is the mass of the lowest
$1^-$ glueball state. The dots denote the contributions of heavier states. 
Thus, according to Eqs. (3.1) and (3.6), the coefficient functions in the IR limit behave like

\begin{equation}                           
F(q^2) = {1 \over M^2_B} + {C \over M^2_{gl}} + O(q^2), \qquad
G(q^2) = - 1 +  O(q^2), \qquad  q^2 \rightarrow 0.    
\end{equation}
At the same time according to Eqs. (3.3), (3.5) and (3.6) their truly nonperturbative counterparts behave like
$G^{NP}(q^2) = q^2 F^{NP}(q^2) =  - 1 + O(q^2), \ q^2 \rightarrow 0$,    
i.e., in the same way as $G(q^2)$ in Eq. (3.7).

\section{Vacuum structure in the Abelian Higgs model}

Let us calculate the truly nonperturbative VED in the Abelian
Higgs model described in the preceding section.                                
It is instructive to start from the case when there are no string contributions
into the structure functions $F(q^2)$ and $G(q^2)$. Then their truly nonperturbative parts are given in Eqs. (3.4). It is convenient to factorize the scale dependence of VED by introducing dimensionless variables and parameters in accordance with B-scheme (2.19) with $\Lambda_{NP} = M_B$.
In this case, the gluon form factors (structure functions) obviously becomes   
the functions of $z$ only, $q^2F^{NP}(q^2)= G^{NP}(q^2)= - (1 / 1+z)$.
The truly nonperturbative VED (2.22) becomes

\begin{equation}
\bar \Omega_g (z_0) = { 1 \over q_0^4} \epsilon_g^{np}(z_0) = 
 - {1 \over \pi^2} z_0^{-2}   
\int_0^{z_0} dz \ z \left[ {2 \over 1 + z} + \ln \left( {-1 + z \over 1 + z } \right) \right].
\end{equation}
Easily integrating Eq. (4.1), one obtains 

\begin{equation}
\bar \Omega_g (z_0) = - {1 \over 2 \pi^2}  z_0^{-2} \left[ 2 z_0 - 4 \ln (1 + z_0) + \ln \left({1 + z_0 \over 1 - z_0} \right) + z_0^2 \ln \left( {- 1 + z_0 \over 1 + z_0} \right) \right].
\end{equation}
From this expression it is almost obvious that the effective potential will have imaginary part at any finite value of the soft cutoff, which is a direct manifestation of the vacuum instability [29]. Asymptotics of the effective potential (4.2) to-leading order are

\begin{eqnarray}
\bar \Omega_g (z_0)_{z_0 \rightarrow 0} &\sim & - {1 \over 2 \pi^2} \ln (-1), \nonumber\\
\bar \Omega_g (z_0)_{z_0 \rightarrow \infty} &\sim & {2 \over \pi^2} z_0^{-2} \ln z_0.
\end{eqnarray}
Let us remind, that $z_0 \rightarrow \infty$ is the perturbative limit ($M_B \rightarrow 0$) when the soft cutoff $q_0$ is fixed. The "stationary" condition  
(2.23) now is 
 
\begin{equation}
4 \ln (1 + z_0) - \ln \left({1 + z_0 \over 1 - z_0} \right) = {z_0 \over 1 + z_0} \left( (1 + z_0)^2 + 2 \right).
\end{equation}
It has only trivial solution $z_0 =0$, so the "stationary" state does not exist
in this model.

\section{Conclusions}

In summary, we have formulated a general method how to calculate the truly nonperturbative VED in the axial gauge QCD quantum models of its 
ground state using the effective potential approach for composite operators. It
is defined
as integrated out the truly nonperturbative part of the full gluon propagator over the deep IR region (soft momentum region). The nontrivial minimization procedure which can be done only by the two different ways (leading however to the 
same numerical value (if any) of VED) 
makes it possible to determine the value of the soft cutoff in terms of the corresponding nonperturbative scale parameter which is inevitably presented in any
nonperturbative model for the full gluon propagator. If the chosen Ansatz for the full gluon propagator is a realistic one, then our method uniquely determines the truly nonperturbative VED, which is always finite, automatically negative
and it has no imaginary part (see, for example our previous publications [23,28]). Here we illustrate it by considering the Abelian Higgs model of the dual QCD ground state. 
The quantum part of VED (4.2) always contains imaginary part. Thus, the vacuum 
of the Abelian Higgs model without string contributions is unstable indeed. Whether the string
contributions can cure this fundamental problem or not is beyond the scope of this letter and is left for consideration elsewhere. 
The vacuum instability of the Abelian Higgs model without string contributions 
will be recovered, of course, within A-scheme (2.16) as well. Nothing depends  
on how one introduces the scale dependence by chosing different scale parameters.                                                                   

Comparing Eqs. (2.9) and (2.15), a prescription to obtain
the relevant expression for the truly nonperturbative VED can
be derived. Indeed, for this purpose in Eq. (2.9) the replacement $q^2 F(q^2) +
G(q^2) \rightarrow 1 + q^2 F^{NP}(q^2) + G^{NP}(q^2)$ should be done.          
Also the soft cutoff $q_0^2$ on the upper
limit should be introduced. Now it looks like the UV cutoff, but nevertheless let us underline once more that it separates the deep IR region from the perturbative one, which includes the IM region as well. It can not be arbitrary large 
as the UV cutoff is, by definition. As far as one chooses Ansatz for the full gluon propagator, the separation
"NP versus PT" in Eq. (2.10) is exact because of the definition (2.11). The separation "soft versus hard" momenta is also exact because of the above-mentioned
minimization procedure. Thus the proposed determination of the truly nonperturbative VED is uniquely defined. It is possible to minimize the effective potential at a fixed scale 
(2.17) with respect to the physically meaningful parameter. When it is zero, the perturbative phase only survives in all quantum models of the QCD ground state. Equivalently, we can minimize the auxiliarly effective potential (2.22) as 
a function of the soft cutoff itself. As was underlined above, both methods lead to the same numerical value for the truly nonperturbative VED.

There is no general method to formulate in order to calculate the confining quark quantum contribution into the total VED since this contribution depends heavily on the particular solutions to the quark SD equation. 
If it is correctly calculated then it is of opposite sign to the nonperturbative gluon part and it is one order of magnitude less as well (see, for example  our recent papers [19,23,28]). Concluding, let us note that the generalization
of our method on different noncovariant gauges [26,30] is straightforward. Let 
us underline that our method is not a solution to the above-mentioned fundamental badly divergent problem of VED. However, it is a general one and can be applied to any nontrivial QCD quantum vacuum models in order to extract the finite 
part of the truly nonperturbative VED in a self-consistent way. In particular, 
it can serve as a test of different axial gauge QCD quantum as well as classical vacuum models since our method provides an exact criterion for the separation
"stable versus unstable vacua". 


\acknowledgments

One of the authors (V.G.) is grateful to M. Polikarpov, M. Chernodub, A. Ivanov
and especially to V.I. Zakharov for useful and critical discussions which led finally the authors to the formulation of a general method presented here. He also would like to thank H. Toki for many discussions on this subject during his 
stay at RCNP, Osaka University. This work was supported by the OTKA grants 
No: T016743 and T029440.

\vfill

\eject


\begin{references}
\bibitem{1}
   P. van Baal (Ed.), Confinement, Duality, and Nonperturbative Aspects of QCD,
NATO ASI Series B: Physics, vol. 368; \\
   K-I. Aoki, O. Miymura, T. Suzuki (Eds.), Non-Perturbative QCD, Structure of 
the QCD Vacuum, Prog. Theor. Phys. Suppl. {\bf 131}, 1 (1998)   
\bibitem{2}
   T. DeGrand, A. Hasenfratz, T.G. Kovacs, Nucl. Phys. {\bf B505}, 417 (1997); 
\\
   D.A. Smith, M.J. Teper, Phys. Rev. D {\bf 58}, 014505 (1998)
\bibitem{3}
   V.N. Gribov, Eur. Phys. Jour. C {\bf 10}, 91 (1999)
\bibitem{4}
   J. Hosek, G. Ripka, Z. Phys. A {\bf 354}, 177 (1996); \ H.G. Dosch, Prog. Part. Nucl. Phys. {\bf 33}, 121 (1994); \
   Yu.A. Simonov, Phys. Usp. {\bf 166}, 337 (1996)
\bibitem{5}
   T. Schafer, E.V. Shuryak, Rev. Mod. Phys. {\bf 70}, 323 (1998)
\bibitem{6}
   J.M. Cornwall, R. Jackiw, E. Tomboulis, Phys. Rev. D {\bf 10}, 2428 (1974)
\bibitem{7}
   A. Barducci et al., Phys. Rev. D {\bf 38}, 238 (1988)
\bibitem{8}
   R.W. Haymaker, Riv. Nuovo Cim. {\bf 14}, 1-89 (1991); \\
   K. Higashijima, Prog. Theor. Phys. Suppl. {\bf 104}, 1-69 (1991)
\bibitem{9}
   W. Marciano, H. Pagels, Phys. Rep. C {\bf 36}, 137 (1978)
\bibitem{10}
   C.D. Roberts, A.G. Williams, Prog. Part. Nucl. Phys. {\bf 33}, 477 (1994);\\
   V. Gogohia, Gy. Kluge, M. Prisznyak, hep-ph/9509427
\bibitem{11}
   M. Baker, C. Lee, Phys. Rev. D {\bf 15}, 2201 (1977)
\bibitem{12}
   E.G. Eichten, F.L. Feinberg, Phys. Rev. D {\bf 10}, 3254 (1974)
\bibitem{13}
   J.E. Mandula, Phys. Rep. {\bf 315}, 273 (1999)
\bibitem{14}
   M.S. Chanowitz, S. Sharpe, Nucl. Phys. {\bf B222}, 211 (1983)
\bibitem{15}
   R.J. Crewther, Phys. Rev. Lett. {\bf 28}, 1421 (1972); \\
   M.S.  Chanowitz, J. Ellis, Phys. Rev. D {\bf 7}, 2490 (1973); \\
   J.C. Collins, A. Duncan, S.D. Joglecar, Phys. Rev. D {\bf 16}, 438 (1977)
\bibitem{16}
   M.A. Shifman, A.I. Vainshtein, V.I. Zakharov, Nucl. Phys. {\bf B147}, 385 (1979)
\bibitem{17}
   G.A. Christos, Phys. Rep. {\bf 116}, 251 (1984)
\bibitem{18}
   E. Witten, Nucl. Phys. {\bf B156}, 269 (1979); \\
   G. Veneziano, Nucl. Phys. {\bf B159}, 213 (1979)
\bibitem{19}
   V. Gogohia, H. Toki, Phys. Lett. B {\bf 466}, 305 (1999); \\
   V. Gogohia, H. Toki, Phys. Rev. D {\bf 61}, 036006 (2000), see also hep-ph/9908301
\bibitem{20}
   V.A. Novikov, M.A. Shifman, A.I. Vainshtein, V.I. Zakharov, Nucl. Phys. {\bf
B191}, 301 (1981)
\bibitem{21}
   I. Halperin, A. Zhitnitsky, Nucl. Phys. {\bf B539}, 166 (1999)
\bibitem{22}
   G. Gabadadze, Phys. Rev. D {\bf 58}, 094015 (1998)
\bibitem{23}
   V. Gogohia, Gy. Kluge, H. Toki, T. Sakai, Phys. Lett. B {\bf 453}, 281 (1999)
\bibitem{24}
   M.N. Chernodub, M.I. Polikarpov, V.I. Zakharov, hep-ph/9903272
\bibitem{25}
   M. Baker, J.S. Ball, F. Zachariasen, Phys. Rev. D {\bf 37}, 1036 (1988)
\bibitem{26}
   A.I. Alekseev, B.A. Arbuzov, Phys. Atom. Nucl. {\bf 61}, 264 (1998); \\
   A.I. Alekseev, B.A. Arbuzov, Mod. Phys. Lett. A {\bf 13}, 1747 (1998)
\bibitem{27}
   P. Castorina, S.-Y. Pi, Phys. Rev. D {\bf 31}, 411 (1985)
\bibitem{28}
   V. Gogohia, H. Toki, T. Sakai, Gy. Kluge,  Inter. Jour. Mod. Phys. A {\bf 15},  45 (2000), see also hep-ph/9810516
\bibitem{29}
   N.K. Nielsen, P. Olesen, Nucl. Phys. {\bf B144}, 376 (1978)
\bibitem{30}
   G. Leibbrandt, Noncovariant gauges, (Word Scientific, Singapore, 1994)
\end{references}
\end{document}